\newcommand{\be}[0]{\begin{equation}}
\newcommand{\ee}[0]{\end{equation}}
\newcommand{\ba}[0]{\begin{eqnarray}}
\newcommand{\ea}[0]{\end{eqnarray}}

\newcommand\GeV{\,\mbox{GeV}}

\documentclass[preprint,12pt]{elsarticle}

\usepackage{axodraw}
\usepackage{graphicx}
\usepackage{epsfig}
\usepackage{amssymb}

\journal{Elsevier}

\begin{document}

\begin{frontmatter}



\title{$K_{S}$ $\to$ $\gamma \gamma$ at one-loop revisited}


\author{Karim Ghorbani\footnote{k-ghorbani@araku.ac.ir}}

\address{Physics Department, Arak University, 38156, Arak, Iran}
\address{School of Particles and Accelerators, IPM~(Institute for Studies in Theoretical Physics and
Mathematics), P.O.Box 19395-5531, Tehran, Iran}

\begin{abstract}
We have calculated the decay amplitude for the process $K_{S}$ $\to$ $\gamma \gamma$ 
at one loop order in chiral perturbation theory. As a new improvement we have 
included the weak mass term which is only relevant for processes with 
external fields in the final state. This term was ignored in earlier
publications for this decay. We find that the inclusion of $G_{8}^{\prime}$ 
brings the theoretical decay rate into a good agreement with experiment.  
\end{abstract}

\begin{keyword}
rare decay, non-leptonic kaon decay, chiral symmetry\\
PACS numbers: 11.30.Rd, 12.39.Fe, 13.25.Es

\end{keyword}

\end{frontmatter}


\section{Introduction}
\label{int}
The non-leptonic kaon decay $K_{S} \to \gamma \gamma$ 
provides a good testing bed for the effective lagrangian 
method at one loop order. The reason hinges in the fact that 
to first order in perturbation theory, short-distance effects are suppressed 
and the decay amplitude to one-loop order in chiral effective theory 
is free from unknown low energy effective constants (LEC's).
For a good recent review on the weak chiral lagrangian see \cite{DAmbrosioTalk}.
The branching ratio with respect to the $\pi^{+}$ $\pi^{-}$ channel 
is theoretically obtained in~\cite{Goity}. The decay rate is also evaluated 
in an independent work~\cite{DAmbrosio}. The theoretical result at $p^{4}$
gives $BR(K_{S}\to \gamma \gamma) = 2.1\times 10^{-6}$~\cite{Goity,DAmbrosio}.
This finding is in good agreement with experimental measurement 
of NA31 that obtained $BR(K_{S}\to \gamma \gamma) = (2.4 \pm .9)\times 10^{-6}$~\cite{NA31-1995}
and with that of KLOE that measured $BR(K_{S}\to \gamma \gamma) = 
(2.26 \pm .12)\times 10^{-6}$~\cite{KLOE2007} . 
On the other hand, the most recent measurement from NA48, 
obtained $BR(K_{S}\to \gamma \gamma)= 2.71\times 10^{-6}$~\cite{NA48-2002} 
with a total uncertainty of about $3\%$.   
The latter experiment opens up the possibility of a sizable 
correction of order $30\%$ from two-loop effects. In view of this 
observation it is deemed interesting to study the effect 
of the higher order corrections for this decay~\cite{KGH}.   
The leading two-loop divergences for the octet part of the non-leptonic
weak sector are already available in \cite{Buchler}.   

The application of weak lagrangian is also extended to other rare process, 
namely, $K \to \pi \gamma \gamma$ in~\cite{Ecker1987,Truong1993,Kambor1993tv}.  
The decays $K \to 2\pi$ and $K \to 3\pi$ are studied at one loop order 
in many places, e.g. see~\cite{Kambor1992he,Kambor1991ah,Bijnens2002}. 

Given all these, we now turn to the main point which motivates the present work.
In~\cite{Kambor1989tz} it is demonstrated that the weak mass term 
appearing in the lowest order weak lagrangian has no contribution 
to the physical amplitude when there is no external fields in the 
decay or when the interaction does not carry four-momentum.
Moreover, it is explicitly shown in~\cite{Bijnens2002} that these effects can 
be reconstructed from redefining the weak effective constants of order $p^{4}$
in the decay $K \to 3\pi$. 
In fact the presence of both strong and weak effective constants 
of $p^4$ order made this procedure possible. 
This type of relations cannot be generalized to the amplitude at order $p^{4}$
in the presence of external fields as in the case of non-leptonic kaon decay 
to two photons, because there are no tree diagram of next-to-leading order for this process. 
For a detailed discussion on the contribution of the weak mass term
on the $K\to \pi \pi$ amplitude we suggest~\cite{BijnensG8p} and references therein. 
We therefore have recalculated the one loop order amplitude for the decay $K_{S} \to \gamma \gamma$ 
and have also taken into account the weak mass term.
The only weak effective constants involved
at order $p^4$ are $G_{8}$, $G_{27}$ and $G_{8}^{\prime}$.
It is easy then to see that the $G_{8}^{\prime}$ effect has 
an essential contribution at order $p^{4}$ to the decay
and cannot be disregarded.

The organization of this letter is as follows. We provide a 
brief introduction to the weak and strong chiral lagrangian 
at leading order in Section 2. In Section $3$ the kinematics is described.
Sections $4$ and $5$ are devoted to our analytical and numerical results 
respectively. Finally we conclude in the last section.

\section{The ChPT Lagrangians}
\label{chpt}
We apply effective lagrangians in order to study the low 
energy dynamics of the strong and weak interactions. 
The lagrangian we use is the lowest order chiral lagrangian.
The expansion parameter is in terms of external momentum $"p"$ 
and quark masses, $"m_{q}"$. 
Quark masses are counted of order $p^2$ due to the lowest 
order mass relation $m_{\pi}^2 = B_{0}(m_{u}+m_{d})$.
Here we only present the leading order strong and weak chiral lagrangian.
The leading order lagrangian which is of order $p^2$, 
assumes the form
\be
{\cal L}_{2} = {\cal L}_{S2} + {\cal L}_{W2}. 
\ee 
${\cal L}_{S2}$ refers to the strong sector with $\Delta S = 0$ 
and ${\cal L}_{W2}$ stands for the weak part with $\Delta S = \pm 1$. 
For the strong part we use~\cite{GL1}
\be
{\cal L}_{S2} = \frac{F_{0}^{2}}{4} \langle u_{\mu} u^{\mu}+ \chi_{+} \rangle,
\ee
where $F_{0}$ is the pion decay constant at chiral limit and we define the 
matrices $u^{\mu}$ and $\chi_{\pm}$ as following

\ba
u_{\mu} = i u^{\dag}D_{\mu}U u^{\dag} = u_{\mu}^{\dag}  \,, \quad u^{2} = U, 
\nonumber\\
\chi_{\pm} = u^{\dag} \chi u^{\dag} \pm u\chi^{\dag}u. 
\ea
The matrix $U \in SU(3)$ contains 
the octet of light pseudo-scalar mesons with its exponential
representation given in terms of meson fields 
matrix as 
\be
U(\phi) = \exp(i \sqrt{2} \phi/F_0)\,,
\ee
where
\ba
\phi (x) 
 = \, \left( \begin{array}{ccc}
\displaystyle\frac{ \pi_3}{ \sqrt 2} \, + \, \frac{ \eta_8}{ \sqrt 6}
 & \pi^+ & K^+ \\
\pi^- &\displaystyle - \frac{\pi_3}{\sqrt 2} \, + \, \frac{ \eta_8}
{\sqrt 6}    & K^0 \\
K^- & \bar K^0 &\displaystyle - \frac{ 2 \, \eta_8}{\sqrt 6}
\end{array}  \right) .
\ea
We use the method of external fields discussed in~\cite{GL1}. 
The external fields are then defined through the covariant derivatives as
\ba
D_{\mu} U = \partial_{\mu} U - i r_{\mu}U +iUl_{\mu}.
\ea
The right-handed and left-handed external fields are expressed 
by $r_{\mu}$ and $l_{\mu}$ respectively. In the present work we set
\ba
r_{\mu} = l_{\mu}  
 = e~A_{\mu} \left( \begin{array}{ccc}
\displaystyle 2/3 &   \\
    &\displaystyle -1/3 &  \\
 &   &\displaystyle -1/3
\end{array}  \right) .
\ea
The electron charge is denoted by $e$ and $A_{\mu}$ is the classical photon field.
The Hermitian $3\times3$ matrix $\chi$ involves the scalar (s) and pseudo-scalar external 
densities and is given by $\chi = 2B_{0}(s+ip)$. 
The constant $B_{0}$ is related to the pion decay 
constant and quark condensate. For our purpose it suffices to write 
\ba
\chi 
 = 2B_{0}\, \left( \begin{array}{ccc}
\displaystyle m_{u} &   \\
    &\displaystyle m_{d} &  \\
 &   &\displaystyle m_{s}
\end{array}  \right) .
\ea
The $\Delta S = \pm 1$ part of the weak effective lagrangian contains both  
the $\Delta I = 1/2$ piece and the $\Delta I = 3/2$ transition and has 
the form~\cite{Cronin1967}
\ba
{\cal L}_{W2} = C F_{0}^{4} \Big[ G_{8} \langle \Delta_{32} u_{\mu} u^{\mu} \rangle 
+G^{\prime}_{8} \langle \Delta_{32}\chi_{+} \rangle
\nonumber \\
+ G_{27} t^{ij,kl} \langle \Delta_{ij}u_{\mu}\rangle \langle \Delta_{kl}u^{\mu} \rangle \Big]
+ h.c,
\ea
where the coefficient $C$ is defined as 
\ba
C = -\frac{3}{5}\frac{G_{F}}{\sqrt{2}}~V_{ud}V_{us}^{*}\approx -1.09 \times 10^{-6}~\GeV^{-2}
\ea 
and the matrix $\Delta_{ij}$ is given by 

\ba
\Delta_{ij} = u \lambda_{ij} u^{\dag}  \,, \quad (\lambda_{ij})_{ab} = \delta_{ia}\delta_{jb}.  
\ea
The nonzero components of the tensor $t^{ij,kl}$ are 
\ba
t^{21,13} = t^{13,21} = \frac{1}{3} &\,,& \quad  t^{22,23} = t^{23,22} = -\frac{1}{6}
\nonumber\\
t^{23,33} = t^{33,23} = -\frac{1}{6} &\,,& \quad  t^{23,11} = t^{11,23} = \frac{1}{3}.
\ea 
The constant $F_{0}$ is the pion decay constant at chiral limit.

\section{Kinematics}
\label{kin}
The decay amplitude of $K_{S} \to \gamma \gamma$ with
the following momentum assignment
\ba
K_{S}(p)  \to   \gamma (k_{1}) \gamma (k_{2}),   
\ea
has the form
\ba
A( K_{S} \to \gamma \gamma ) =  M_{\mu\nu}(k_{1},k_{2}) \hspace{.1cm} 
{\epsilon_{1}}^{\mu} (k_{1}) \hspace{.1cm} {\epsilon_{2}}^{\nu}(k_{2}),
\ea
where ${\epsilon_{1}}^{\mu}$ and ${\epsilon_{2}}^{\nu}$ are the 
polarization four-vectors of the outgoing photons
carrying momentum $k_{1}$ and $k_{2}$ respectively.
Due to the gauge invariance, Lorentz symmetry and Bose symmetry, $M_{\mu\nu}(k_{1},k_{2})$ 
takes on the specific form
\ba
M_{\mu\nu}(k_{1},k_{2}) = F(p^2)\hspace{.1cm}(  k_{1\nu} k_{2\mu} - 
k_{1}.k_{2} \hspace{.1cm} g_{\mu\nu} ).
\ea
Where $p = k_{1}+k_{2}$ and $k_1^2 = k_2^2 = 0$ for photons with on-shell masses. 
The decay width for a decay with two particles in the final state reads
\ba
\Gamma(K_{S} \to \gamma \gamma) = \frac{1}{16\pi m_{K}} |F(p^2 = m_{K}^2)|^2.
\ea

\section{Analytical Results}
\label{ARes}
\begin{figure}
\begin{center}
\includegraphics[width=.48\textwidth]{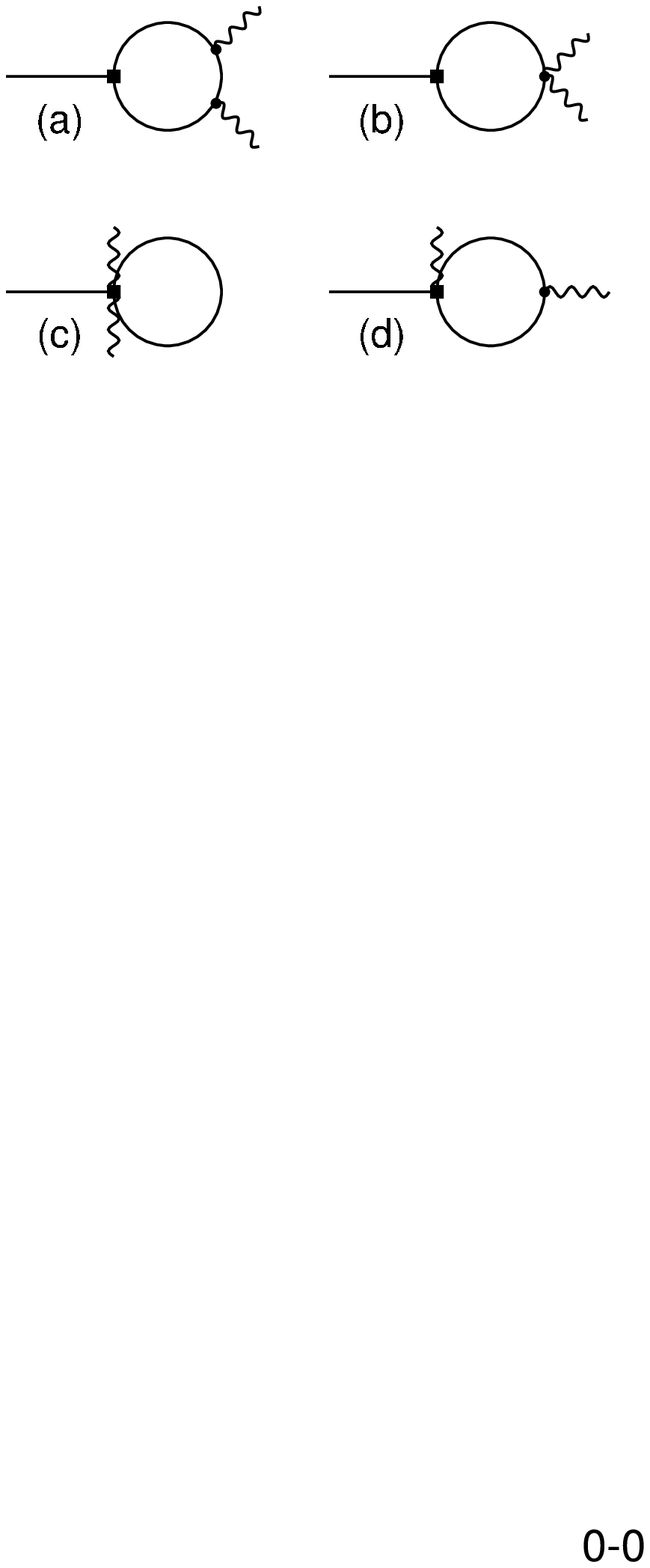}
\end{center}
\caption{Feynman diagrams of order $p^4$. 
A filled circle is the strong vertex from ${\cal L}_{S2}$
and a filled square is the weak vertex from ${\cal L}_{W2}$.
Solid lines represent the pseudo-scalar meson particles and wavy lines stand for photons.}
\label{loop4}
\end{figure}

The decay amplitude gets no tree-level contribution of order $p^{2}$ and $p^{4}$.
This is because all the particles involved here are neutral particles. Thus,
the leading non-zero part of the amplitude originates from loop diagrams constructed 
out of strong and weak lagrangians of order $p^{2}$. The relevant Feynman diagrams
for this decay is depicted in Fig.~\ref{loop4}. Since tree diagrams are absent 
here we therefore expect that the sum of all the Feynman diagrams ends up 
finite, i.e. all infinities from loop integrals cancel.
This is indeed proven by our explicit calculation. We present our 
result in a form that full agreement with the earlier results given in \cite{Goity,DAmbrosio} 
can be simply understood besides an extra term in our expression followed by the 
coupling constant $G_{8}^{\prime}$ which is new. 
We use the lowest order relations $m_{\pi}^2 = B_{0} (m_u + m_d)$ 
and $m_{K}^2 = B_{0} (m_{s} + m_{u})$
to replace quark masses with the mesons masses.  
The following analytical result is achieved  
\ba
\label{formula1}
F(p^2) &=& \frac{C(G_8+G_{27}-4/3G_8^{\prime})\alpha_{em}F_{\pi}}{2\pi} \Big(\frac{p^2-m_{\pi}^2}{p^2} \Big)
\nonumber \\&&
\Big(1+\frac{m_{\pi}^2}{p^2}\log^2 \Big( \frac{\sqrt{1-4m_{\pi}^2/p^2}-1}{\sqrt{1-4m_{\pi}^2/p^2}+1} \Big) \Big) 
\nonumber \\&&
-(\pi \to K)
\ea
using the program FORM~\cite{Form3}.
It should be noticed that for $p^{2} = m_{K}^2$, the contribution of $G_{8}^{\prime}$ reduces 
the magnitude of $F(m_{K}^2)$.

\section{Numerical results}
In this section we estimate the decay rate for two sets of inputs.
We use the physical values instead of lowest order values for masses 
and pion decay constants in doing numerics. This is because the 
difference stands on the higher chiral order.
We use then $F_{\pi} = 0.092$~GeV for the pion decay constant, $m_{\pi} = 0.1395$~GeV
and $m_{K} = 0.4936$~GeV for the charged pion and kaon mass respectively.
The quantities $G_{8}$ and $G_{27}$ are determined at one-loop ChPT order 
by performing direct fit to the experimental 
data in decays $K \to \pi \pi \pi$ and $K \to \pi \pi$.
For detailed discussions we refer the reader to ~\cite{Bijnens2002}.
The values of $G_{8}$ and $G_{27}$ which are presented in~\cite{Bijnens2002} read 
\ba
\label{Gs1}
G_{8} = 5.49\pm 0.02 \qquad  G_{27} = 0.392\pm 0.002
\ea 
We quote these values as "Set 1" hereafter. 
There is also another estimate for these quantities based on a hadronic model including
a $Q_{2}$ penguin-like contribution that obtained~\cite{Bijnens1998} 
\ba
\label{Gs2}
G_{8} = 6.0\pm 1.7 \qquad  G_{27} = 0.35\pm 0.15. 
\ea 
We quote these values as "Set 2" hereafter. 
This set of values inherits large uncertainties,
however, we shall use them for the sake of comparison.    
There is one study concerning the determination of the quantity $G_{8}^{\prime}$ 
employing a hadronic model at next to leading order in large $N_{c}$ ~\cite{Bijnens2000}.
They obtained 
\ba
\label{Gp}
G_{8}^{\prime} = 0.9\pm 0.1
\ea
We set $G_{8}^{\prime} =0$ in Eq.~\ref{formula1} and use the values of 
$G_{8}$ and $G_{27}$ given in Eq.~\ref{Gs1} and Eq.~\ref{Gs2} to obtain the theoretical decay rate 
\ba
\Gamma(K_{S} \to \gamma \gamma)_{th} &=& (2.879\pm 0.02 )\times 10^{-20}~GeV,  \qquad  Set~1
\nonumber\\&&
                                      (3.401\pm 1.828)\times 10^{-20}~GeV.    \qquad  Set~2
\ea
The estimated errors are due to the uncertainties in $G_{8}$ and $G_{27}$.
As we expect from the formula in Eq.~\ref{formula1}, a non-zero value for $G_{8}^{\prime}$
lowers the value of the decay rate such that we obtain   
\ba
\label{rate}
\Gamma(K_{S} \to \gamma \gamma)_{th} &=& (1.817\pm 0.165 )  \times 10^{-20}~GeV,  \qquad  Set~1
\nonumber\\&&
                                       (2.237\pm 1.494 ) \times 10^{-20}~GeV.    \qquad  Set~2
\ea
We compare our theoretical result with the averaged experimental measurements provided in \cite{PDG2008}
\ba
\Gamma(K_{S} \to \gamma \gamma)_{exp} = (2.115\pm 0.136) \times 10^{-20} ~GeV.  \qquad  
\ea
Using the values for the theoretical decay rate given in Eq.~\ref{rate}
and the experimental total decay rate $\Gamma(K_S)_{exp} = (7.385\pm 0.026)\times 10^{-15}$~\cite{PDG2008}~GeV,
it is possible to obtain the branching ratio as    
\ba
BR(K_{S} \to \gamma \gamma)_{th} &=& (2.46\pm 0.22) \times 10^{-6},  \qquad  Set~1
\nonumber\\&&
                                      (3.02\pm 2.02) \times 10^{-6}.  \qquad  Set~2
\ea 
The experimental branching ratio provided in~\cite{PDG2008} reads
\ba
BR(K_{S} \to \gamma \gamma)_{exp} = (2.63\pm 0.17) \times 10^{-6}.  
\ea

\section{Conclusion}
In this letter we have recalculated the $K_{S} \to \gamma \gamma$ at one-loop order and
in addition have added a contribution due to the weak mass term 
in the leading order of weak action which had been ignored in the previous works.
We obtained the full result for the decay rate at one-loop order. There are three effective
constants involved, namely, $G_{8}$, $G_{27}$ and $G_{8}^{\prime}$. We have evaluated the 
decay rate using two sets of inputs for $G_{8}$ and $G_{27}$ and 
one single value for $G_{8}^\prime$. It is realized that the effect of the weak 
mass term is important and in order for a one-loop result to reach the 
corresponding experimental data it is necessary to take it into account.

\label{con}

\section{Acknowledgments}
\label{Ack}
I would like to thank Gerhard Ecker for fruitful discussions.
I am also grateful to Hans Bijnens for helpful discussion and 
careful reading of the manuscript. This work is supported in part by IPM.


\end{document}